\documentclass[prl,twocolumn,showpacs,superscriptaddress, floatfix,reprint]{revtex4-1}

\usepackage{graphicx}
\usepackage{color}
\usepackage{amssymb,amsmath}
\usepackage{hyperref}
\usepackage{bm}
\usepackage{pdfpages}
\usepackage{pgffor}

\newcommand{\op}[1]{\hat #1}

\newcommand{\be}{\begin{equation}}
\newcommand{\ee}{\end{equation}}

\makeatletter
\AtBeginDocument{\let\LS@rot\@undefined}
\makeatother

\begin{document}
\title{Few vs many-body physics of an impurity immersed in a  superfluid of spin 1/2 attractive fermions}

\author{M. Pierce}%
\affiliation{Laboratoire Kastler Brossel, ENS-Universit\'e PSL, CNRS, Sorbonne Universit\'e, Coll\`ege de France.
}%

\author{X. Leyronas}
\affiliation{Laboratoire de physique de l'Ecole normale sup\'erieure, ENS, Universit\'e PSL, CNRS,
	Sorbonne Universit{\'e}, Universit\'e Paris-Diderot, Sorbonne Paris Cit\'e, Paris, France.
}%

\author{F. Chevy$^{1}$}%

\begin{abstract}
In this article we investigate the properties of an impurity immersed in a superfluid of strongly correlated spin 1/2 fermions. For resonant interactions, we first relate  the stability diagram of dimer and trimer states to the three-body problem for an impurity interacting with a pair of fermions. Then we calculate the beyond-mean-field corrections to the energy of a weakly interacting impurity. We show that these corrections are divergent and have to be regularized by properly accounting for three-body physics in the problem.
\end{abstract}

\maketitle

The physics of an impurity immersed in a many-body ensemble is one of the simplest although non-trivial paradigms in many-body physics. One of the first examples of such a system is the polaron problem which was introduced  by Landau and Pekar \cite{landau1948effective} to describe the interaction of an electron with the acoustic excitations of a surrounding crystal. Likewise,  in magnetic compounds Kondo's Effect arises from the interaction of magnetic impurities with the background Fermi sea \cite{kondo1964resistance,anderson1961localized}. Similar situations  occur in high-energy physics, e.g. in neutron stars  to interpret the interaction of a proton with a superfluid of neutrons  \cite{zuo20041s0}, or in quantum chromodynamics where the so-called Polyakov loop describes the properties of a test color charge immersed in a hot gluonic medium  \cite{fukushima2017polyakov}. Finally, impurity problems can be used as prototypes for more complex many body-systems \cite{levinsen2017universality}, as illustrated by the dynamical mean-field theory \cite{Georges96dynamical}.

The recent advent of strongly correlated quantum gases permitted by the control of interactions in these systems have opened a new research avenue for the physics of impurities \cite{bloch2008many,zwerger2012BCSBEC,Chevy2010Unitary,massignan2014polarons}. Experiments on strongly polarized Fermi gases \cite{zwierlein2006fsi,partridge2006pap,nascimbene2009pol} were interpreted by the introduction of the so-called Fermi polarons, a quasi-particle describing the properties of an impurity immersed in an ensemble of spin-polarized fermions and dressed by a cloud of particle-hole excitations of the surrounding Fermi Sea \cite{chevy2006upa,lobo2006nsp,prokof'ev08fpb}. More recently, the physics of Bose polarons (impurities immersed in a Bose-Einstein condensate) was explored using radio-frequency spectroscopy \cite{jorgensen2016observation,hu2016bose}. Contrary to the Fermi polaron, this system is subject to an Efimov effect \cite{efimov73energy} and three-body interactions play an important role in the strongly correlated regime \cite{levinsen2015impurity}.  Finally recent experiments on dual superfluids have raised the question of the behaviour of an impurity immersed in a superfluid of spin 1/2 fermions \cite{Ferrier2014Mixture,roy2017two,yao2016observation}. In these experiments, the polaron was weakly coupled to the background superfluid and the interaction could be accurately modeled within mean-field approximation.  Further theoretical works explored the strongly coupled regime using mean-field theory to describe the fermionic superfluid \cite{nishida2015polaronic,yi2015polarons}. They highlighted the role of Efimov physics in the phase diagram of the system and as a consequence some results were plagued by unphysical ultraviolet divergences. In this letter we address this problem without making any assumption on the properties of the superfluid component. We calculate the first beyond-mean-field corrections to the energy of the polaron and we show that the logarithmic divergence arising from three-body physics can be cured within an effective field theory approach introduced previously in the study of beyond mean-field corrections in Bose gases \cite{braaten1999quantum,braaten2002dilute}.

\begin{figure}
    \centering
    \includegraphics[width=\columnwidth]{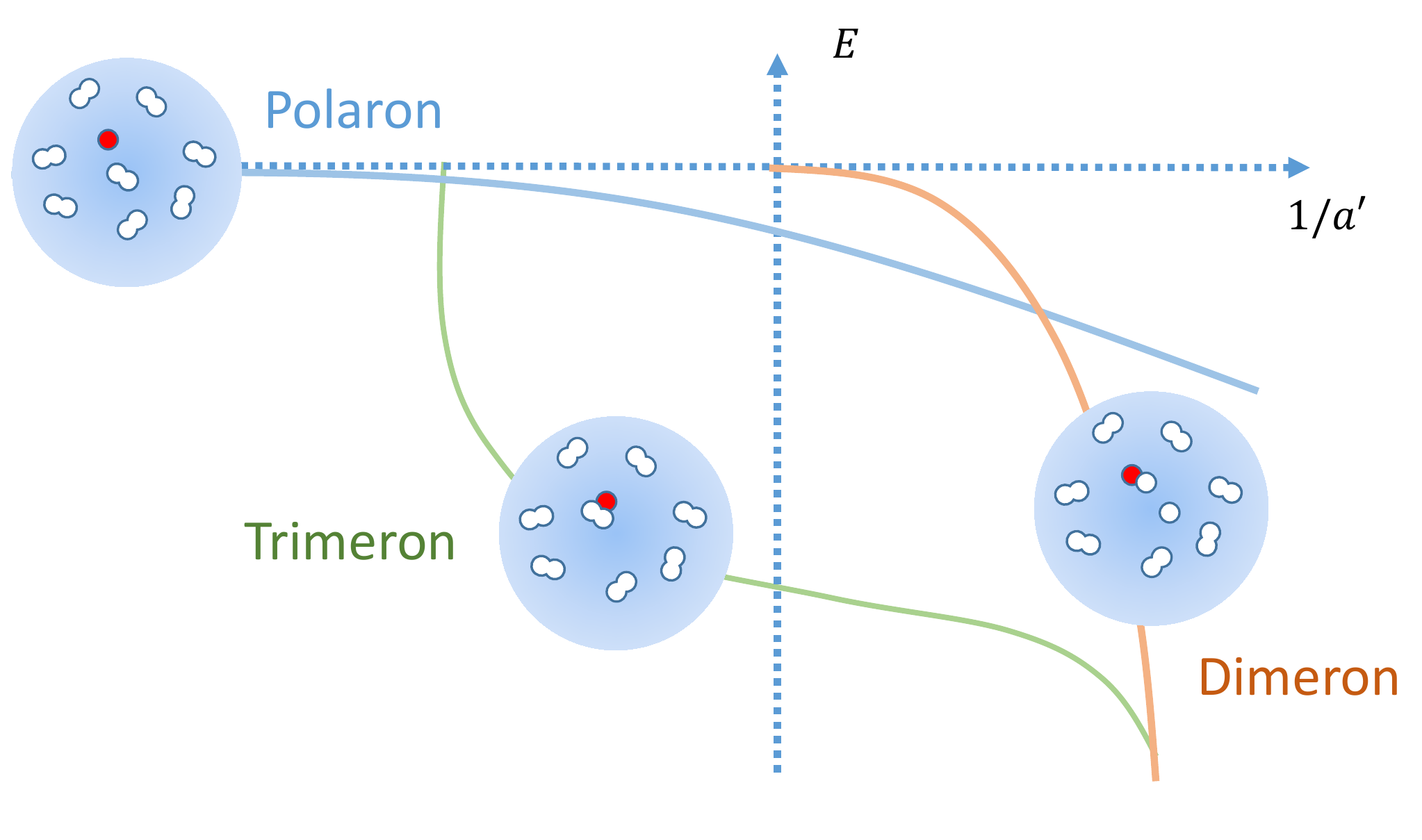}
    \caption{Sketch  of the energy branches of an impurity (red dot) immersed in an ensemble of Cooper-paired fermions when the impurity/fermion scattering length $a$' is varied.  Blue: polaron branch; green: ground-state Efimov trimer branch; orange: dimer branch.  According to mean-field calculations \cite{nishida2015polaronic,yi2015polarons} the  polaron/trimer transition  corresponds to a smooth avoided crossing between the two branches.}
    \label{fig:Fig1}
\end{figure}

Qualitatively speaking, the phase diagram of the impurity can be decomposed in three different regions when the strength of the impurity/fermion interaction is varied (see Fig. \ref{fig:Fig1}). For a weak attraction, the impurity can be described as a polaronic quasi-particle. When attraction is increased, the impurity binds to an existing Cooper pair and the polaronic branch connects to the resonant Efimov trimer states. Earlier variational calculations suggest that the transition between the polaron and trimeron states is a smooth crossover \cite{nishida2015polaronic,yi2015polarons}. Finally, in the strongly attractive regime, impurity/fermion attraction overcomes Cooper pairing leading to a dimeron state describing an impurity/fermion dimer immersed in a fermionic superfluid medium.

\begin{figure}
    \centering
        \includegraphics[width=\columnwidth]{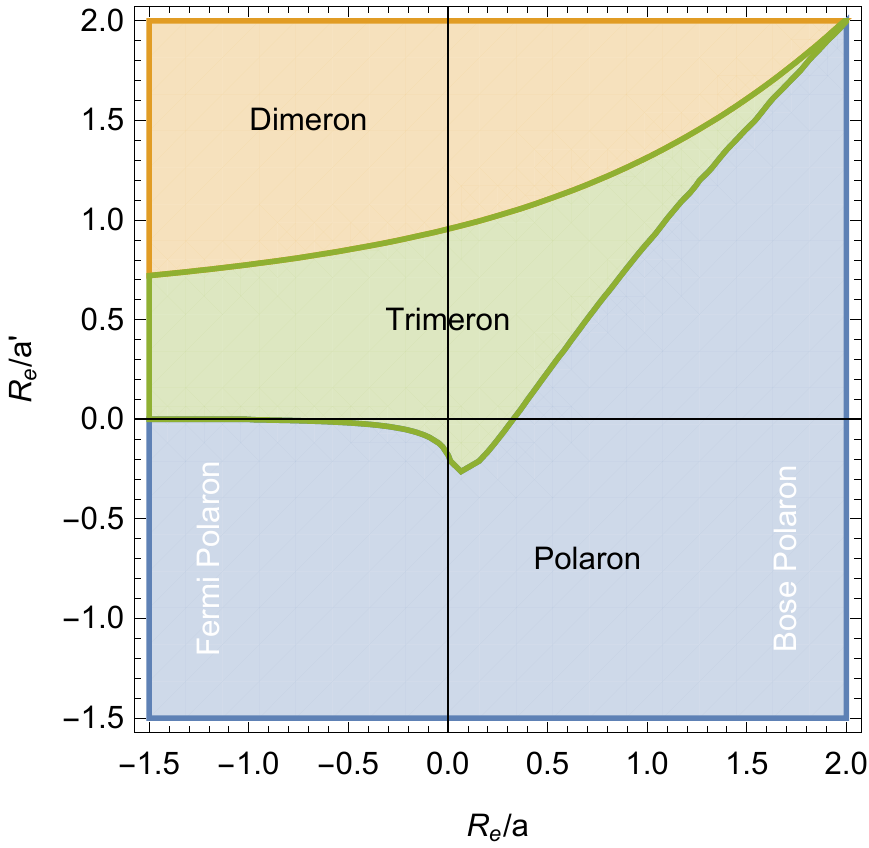}
    \caption{Stability diagram of the polaron, dimeron and   trimeron states with the effective range for $m_{\rm f}=m_{\rm i}$ using a coupled channel model (see text and \cite{SuppMat}). Variational approaches based on a mean-field description of the background superfluid suggest that the polaron/trimeron transition is a crossover \cite{nishida2015polaronic,yi2015polarons}.  }
    \label{fig:Fig2}
\end{figure}

Since the size of the ground-state Efimov trimer is typically much smaller than the interparticle spacing, its binding energy is  much larger than the Fermi energy of the fermionic superfluid. As a consequence, except when the Efimov trimer becomes resonant with the atomic continuum, the internal structure of the trimer is only weakly affected by the many-body environment. A first insight on the phase diagram of the system can thus be obtained from the study of the three-body problem to determine the stability domain of the Efimov trimers with respect to the free-atom and atom-dimer continuum. In this pursuit, we use a two-channel model similar to the one presented in e.g. \cite{gogolin2008analytical} in the case of the bosonic Efimov problem \cite{SuppMat}. For the sake of simplicity, we assume here that the masses of the fermions and the impurity are the same, and that the impurity interacts the same way  with both spin states of the fermionic ensemble (these assumptions are well satisified in the experiments reported in \cite{Ferrier2014Mixture}). The properties of the system are therefore characterized by three different length scales: the fermion-fermion scattering length ($a$), the fermion-impurity scattering length ($a'$) and the effective range of the interaction potential ($R_e$) \footnote{Note that we define $R_e$ as in \cite{gogolin2008analytical} that is positive for a two-channel model and whose sign is opposite to the traditional definition of the effective range.}. The corresponding phase diagram is displayed in Fig. \ref{fig:Fig2}. When the fermion/fermion interaction strength is varied, the superfluid explores the BEC-BCS crossover \cite{zwerger2012BCSBEC} that connects the weakly attractive regime ($a\rightarrow 0^-$) where the fermions form loosely bound Cooper pairs described by BCS (Bardeen-Cooper-Schrieffer) theory, to the strongly attractive limit ($a\rightarrow 0^+$) where they form a Bose-Einstein Condensate (BEC) of deeply bound dimers. As a consequence, the polaronic state smoothly evolves from a Fermi polaron (an impurity immersed in a non-interacting Fermi sea) to a Bose polaron (an impurity immersed in a BEC of dimers). The trimeron stability region is obtained by a numerical resolution of the three-body problem \cite{SuppMat}. The polaron-dimeron frontier simply corresponds to a competition between fermion/fermion and fermion/impurity pairings.

Since the trimeron and dimeron regimes are dominated by few-body physics, we now focus on the energy of the polaronic branch that is more strongly affected by the presence of the superfluid. We furthermore assume that the impurity-fermion interaction can be treated perturbatively.

Consider thus an impurity of mass $m_{\rm i}$  immersed in a bath of spin 1/2 fermions of mass $m_{\rm f}$.  We write the Hamiltonian of the system as

\be
\op H=\op H_{\rm imp}+\op H_{\rm mb}+\op H_{\rm int},
\ee
where $\op H_{\rm imp}$ (resp. $\op H_{\rm mb}$) is the Hamiltonian of the impurity (resp.  many-body background) alone, and $\op H_{\rm int}$ describes the interaction between the two subsystems. We label the eigenstates of the impurity by its momentum $\bm q$ and its eigenvalues are $\varepsilon^{(\rm i)}_q=\hbar^2 q^2/2m_{\rm i}$. The eigenstates and eigenvalues of $\op H_{\rm mb}$ are denoted $|\alpha\rangle$ and $E_\alpha$, where by definition $\alpha=0$ corresponds to the ground state of the fermionic superfluid.

 Assuming for simplicity an identical contact interaction between the impurity and each spin component of the many-body ensemble, we write
\be
\op H_{\rm int}=g'_{0}\sum_{\sigma=\uparrow,\downarrow}\int d^3\bm r \op\psi^\dagger_\sigma(\bm r)\op\psi_\sigma(\bm r)\op\phi^\dagger(\bm r)\op\phi(\bm r),
\ee
where $\op\psi_\sigma$ and $\op\phi$ are the field operators for spin $\sigma$ particles of the many-body ensemble and of the impurity respectively.  In this expression, the bare and physical coupling constants $g'_{0}$ and $g'$ are related through
\be
\frac{1}{g'_{0}}=\frac{1}{g'}-\frac{1}{\Omega}\sum_{k<\Lambda}\frac{1}{\varepsilon^{({\rm r})}_{k}},
\ee
where $\Omega$ is the quantization volume, $\Lambda$ is some ultraviolet cutoff and $\epsilon^{({\rm r})}_k=\hbar^2 k^2/2m_{\rm r}$, with $m_{\rm r}$ the impurity-fermion reduced mass.  Assuming that the contact interaction can be treated perturbatively, we have up to second order
\be
g'_{0}=g'+\frac{{g'}^2}{\Omega}\sum_{k<\Lambda}\frac{1}{\varepsilon^{({\rm r})}_{k}}+o({g'}^2).
\ee

Calculating the energy $\Delta E$ of the polaron to that same order, we have
\be
\Delta E_{\rm pert}=g'  n+\frac{{g'}^2n}{\Omega}\sum_{\bm q}\left[\frac{1}{\varepsilon_{q}^{({\rm r})}}-\chi(\bm q,\varepsilon^{({\rm i})}_q)\right].
\label{eq:Born1}
\ee
where $n$ is the particle density in the many-body medium and \be
\chi(\bm q,E)=\frac{1}{N}\sum_{\alpha}\frac{\left|\langle \alpha|\op\rho_{-\bm q}|0\rangle\right|^2}{E_\alpha-E_0+E},
\ee
with $\op\rho_{\bm q}=\sum_\sigma\int d^3\bm r \op\psi_{\sigma}^\dagger(\bm r)\op\psi_{\sigma}(\bm r)e^{i\bm q\cdot\bm r}$.

In the sum, the presence of the two terms allows for a UV cancellation of their $1/q^2$  asymptotic behaviours. Indeed, for large $q$ the eigenstates of the many-body Hamiltonian excited by the translation operator $\op\rho_{\bm q}$ correspond to free-particle excitations of momentum $\bm q$ and energy $\varepsilon^{({\rm f})}_q=\hbar^2 q^2/2m_{\rm f}$. We therefore have

\be
\chi(\bm q,E)\simeq \frac{S(q)}{\varepsilon^{(\rm f)}_q+E},
\label{eq:chi}
\ee
where $S(q)=\sum_\alpha|\langle \alpha|\op\rho_{\bm q}|0\rangle|^2/N$ is the static structure factor of the many-body system. At large momenta, we have  $S(q)=1+C_2/4Nq+...$,
where $C_2$ is Tan's contact parameter of the fermionic system and characterizes its short-range two-body correlations \cite{hu2010static,tan2008large}. From this scaling we see that the UV-divergent $1/q^2$ contributions in Eq. (\ref{eq:Born1}) cancel out. However, the next-to-leading order term in $S(q)$ suggests that this cancellation is not sufficient to regularize the sum that is still log-divergent. This logarithmic behaviour is supported by a directed calculation of $\chi$ using BCS mean-field theory \cite{SuppMat} and is characteristic of a singularity in the three-body problem for particles with contact interactions that was pointed out first by Wu for bosons \cite{wu1959ground} and was more recently investigated in the context of cold atoms (see for instance \cite{braaten1999quantum,tan2008three,hammer2015three,yi2015polarons}).

To get a better insight on the origin of this singularity, we analyze first the  scattering of an impurity with a pair of free fermions. Within Faddeev's formalism \cite{faddeev2013quantum}, the corresponding three-body $T$-matrix is written as a sum of three contributions, $\op T_{i=1,2,3}$ solutions of the set of coupled equations
\be
\left(\begin{array}{c}
\op{T}_1\\
\op{T}_2\\
\op{T}_3
\end{array}
\right)
=
\left(\begin{array}{c}
\op{t}_1\\
\op{t}_2\\
\op{t}_3
\end{array}
\right)
+
\left(\begin{array}{ccc}
0&\op{t}_1&\op{t}_1\\
\op{t}_2&0&\op{t}_2\\
\op{t}_3&\op{t}_3&0
\end{array}
\right)\op{G}_0
\left(\begin{array}{c}
\op{T}_1\\
\op{T}_2\\
\op{T}_3
\end{array}
\right)
\ee
where $\op{G}_0=1/(z-\op{H}_0)$ is the free resolvant operator and $\op{t}_i$ is the two-body $T$-matrix leaving particle $i$ unaffected. The solutions of this equation can be expressed as a series of diagrams where a given two-body $t$-matrix never acts twice in a row and $\op{T}_i$ corresponds to the sum of all diagrams finishing by $\op{t}_i$. Assuming that the impurity is labeled by the index $i=3$ and that its interaction  with the other two atoms ($i=1,2$) is weak, we can expand the solutions of Faddeev's equation with $\op{t}_1$ and $\op{t}_2$. To be consistent with the polaron-energy calculation outlined in previous section, we proceed up to second order in $\op{t}_{1,2}$.

\begin{figure}
\centerline{\includegraphics[width=\columnwidth]{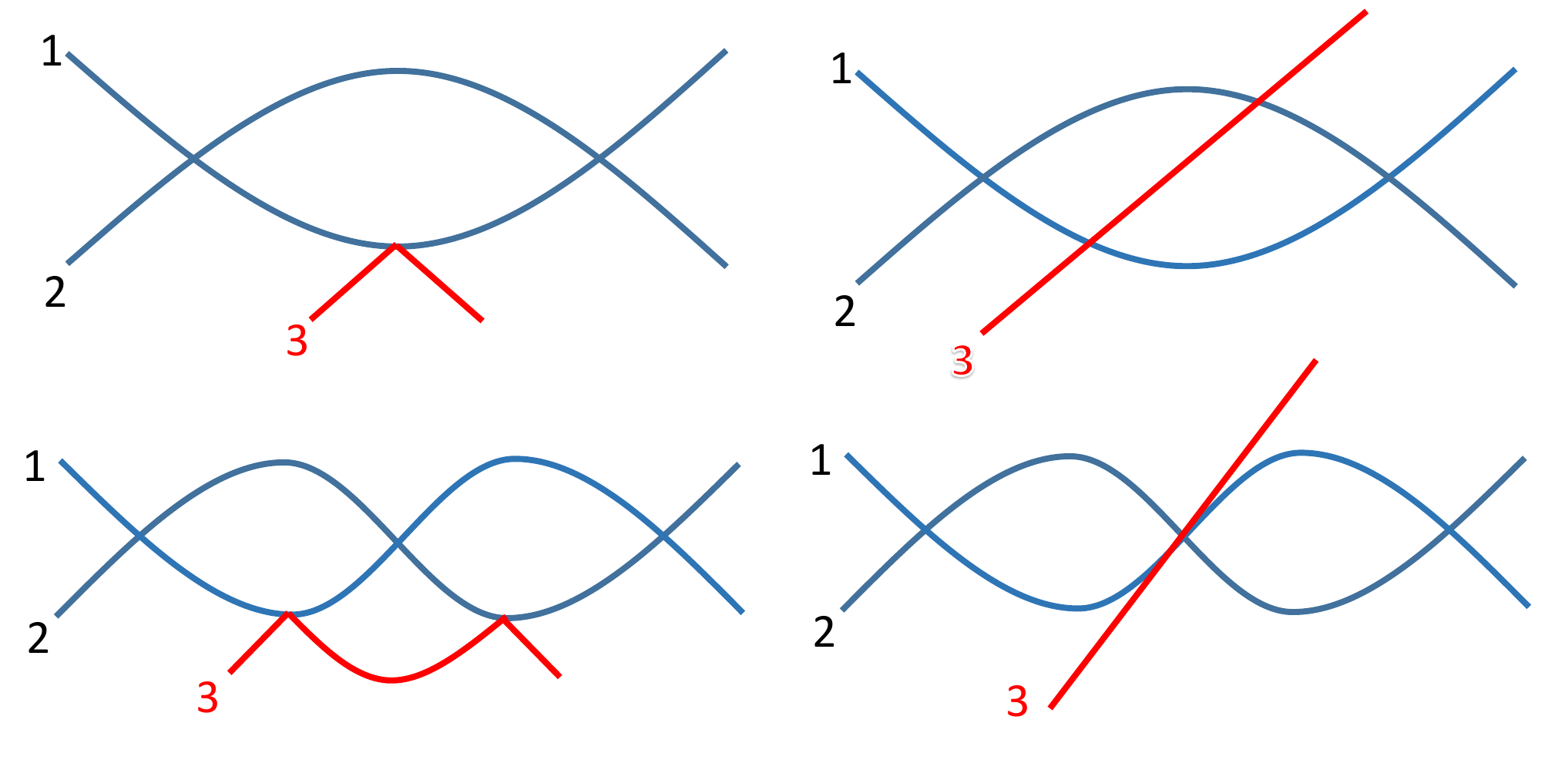}}
\caption{Singular diagrams at second order of Born's approximation. The red line corresponds to the impurity. In Faddeev's expansion, each interaction vertex corresponds to a two-body t-matrix $\op t_{1,2,3}$. In Born's approximation, the $\op t_{1,2}$'s are expanded to second order in $a'$ which leads to a logarithmic UV  divergence of the corresponding terms.}
\label{fig:diagrams}
\end{figure}

When the full 2-body $T$-matrices $ t_{1,2}=g'/\Omega/(1+ia'k+a' R_e k^2)$ are used, all terms of the expansion are finite. However, when treating them within second order Born's approximation (i.e. taking simply $t_{1,2}=g'(1-ika')/\Omega$ and stopping the expansion of the $T$-matrix  at second order in $a$'), some diagrams are logarithmically divergent. The singular diagrams are listed in Fig. \ref{fig:diagrams}: they all start and end with $t_3$ and  their contribution can be written as $(t_3)_{\rm out}\Gamma(t_3)_{\rm in}$. In Born's approximation the sums over inner momenta are divergent and the integrals are therefore dominated by the large-k behaviour of $t_3$ and $G_0$. After a straightforward calculation, we obtain that

\be
\Gamma_{\rm Born}\underset{\Lambda\rightarrow\infty}{\sim} \frac{m_f^3}{\hbar^6} {g'}^2\kappa(\eta=m_b/m_f)\ln(\Lambda)
\ee
with

\be
\begin{split}
\kappa(\eta)=&\frac{\sqrt{\eta^3 (\eta+2)}}{2\pi^3  (\eta+1)^2}-\frac{\eta}{2\pi^3 }\arctan
   \left(\frac{1}{\sqrt{\eta (\eta+2)}}\right)\\
   &-\frac{4 }{\pi^3}  \sqrt{\frac{\eta}{ \eta+2}}\arctan
   \left(\sqrt{\frac{\eta}{\eta+2}}\right)^2
\end{split}
\label{eq:Kappa}
\ee

Since $\Gamma$ is finite when the full two-body physics is taken into account we introduce a three-body characteristic length $R_3$ such that

\be
\Gamma_{\rm Born}-\Gamma_{\rm Faddeev}\underset{\Lambda\rightarrow\infty}{=}\frac{m_f^3}{\hbar^6} {g'}^2\kappa(\eta)\ln(\Lambda R_3)+o(1),
\label{eq:Threebodyparameter}
\ee
where $\Gamma_{\rm Faddeev}$ corresponds to the value of $\Gamma$  obtained by using the full two-body T-matrices $t_{1,2}$ to calculate the first three diagrams of Fig. (\ref{fig:diagrams}). In this perturbative approach, $R_3/a'$ depends on $\eta$ and $R_e/a'$ and can be computed numerically \cite{SuppMat}. Since we work in a regime where the polaron is the ground state and Efimov trimers are absent, we do not have to  use non-perturbative approaches compatible with Efimov physics and leading to a log-periodic dependence of $R_3$ \cite{bedaque1999renormalization}.

Following the effective field theory approach discussed in \cite{hammer2015three}, divergences plaguing Born's expansion can be cured by introducing an explicit three-body interaction described by a Hamiltonian

\be
\op  H_{3b}=g_3(\Lambda)\int d^3\bm r\op\psi_1^\dagger(\bm r)\op\psi_2^\dagger(\bm r)\op\psi_3^\dagger(\bm r)\op\psi_3(\bm r)\op\psi_2(\bm r)\op\psi_1(\bm r)
\ee

The contribution of this three-body interaction to $\Gamma$ corresponds to the fourth diagram of Fig. \ref{fig:diagrams} and yields the following expression

\be
\Gamma_{3\rm b}=g_3(\Lambda)\left(\frac{1}{\Omega}\sum_{k<\Lambda}\frac{1}{2\varepsilon^{(\rm f)}_{k}}\right)^2.
\ee
Using this three-body interaction to cure Born's approximation, we must have $\Gamma_{\rm Born}+\Gamma_{\rm 3b}=\Gamma_{\rm Faddeev}$, hence the following expression for the three-body coupling constant
\be
g_3(\Lambda)\left(\frac{1}{\Omega}\sum_{k<\Lambda}\frac{1}{2\varepsilon^{\rm (f)}_{k}}\right)^2=-\frac{m_f^3}{\hbar^6}{g'}^2\kappa(\eta)\ln(\Lambda R_3)
\label{eq:ThreeBodyCoupling}
\ee

The introduction of the three-body Hamiltonian implies a new contribution to the second-order energy shift (\ref{eq:Born1}). This new term amounts to

\be
\Delta E_{3b}=g_3(\Lambda)\langle 0|\op\psi_1^\dagger(\bm r)\op\psi_2^\dagger(\bm r)\op\psi_2(\bm r)\op\psi_1(\bm r)|0\rangle\ee
Using Eq. (\ref{eq:ThreeBodyCoupling}) as well as the properties of Tan's contact parameter, we obtain after a straightforward  calculation

\be
\Delta E_{3b}=-{g'}^2\kappa(\eta) \frac{m_{\rm f} C_2}{\hbar^2\Omega}\ln(\Lambda R_3).
\label{eq:3bodyEnergy}
\ee

Adding this contribution to Eq. (\ref{eq:Born1}), we obtain for the polaron energy $\Delta E=\Delta E_{\rm pert}+\Delta E_{3b}$:

\be
\begin{split}
\Delta E=g'  n\biggl[ 1+&k_F a' F\left(\frac{1}{k_F a}\right)\\
&-2\pi\frac{m_{\rm f}}{m_{\rm r}}\kappa(\eta)\frac{a'C_2}{N}\ln(k_FR_3)+...\biggr],
\end{split}
\label{eq:Born2}
\ee
with
\be
\begin{split}
F\left(\frac{1}{k_Fa}\right)\underset{\Lambda \rightarrow \infty}{=}\frac{2\pi}{k_F}\biggl[\frac{\hbar^2}{m_{\rm r}}\int_{q<\Lambda}&\frac{d^3\bm q}{(2\pi)^3}\left(\frac{1}{\varepsilon^{\rm (r)}_q}-\chi(q,\varepsilon^{(\rm i)}_q)\right)\\
&-\frac{m_{\rm f}}{m_{\rm r}}\kappa(\eta)\frac{C_2}{N}\ln(\Lambda/k_F)\biggr]
    \end{split}
\label{Eq:MainEq}
\ee

Eq. (\ref{eq:Born2}) and (\ref{Eq:MainEq})  are the main results of this paper. They show that the second order correction of the polaron energy is the sum of two terms: a regular term  characterized by the function $F$ defined by Eq. (\ref{eq:Born2}), as well as a second term, characterized by a logarithmic singularity and proportional to the fermionic contact parameter.

The function $F$ is in general hard to compute exactly but we can obtain its exact asymptotic expression in the BEC and BCS limits. When the fermions of the background ensemble are weakly interacting, we must recover the Fermi-polaron problem (see Fig. \ref{fig:Fig2}). For the mass-balanced case $\eta=1$, we obtain $F(-\infty)=3/2\pi$.
In the strongly attractive limit, the fermionic ensemble behaves as a weakly interacting Bose-Einstein condensate of dimers and the polaron energy takes a general mean-field form $g_{\rm ad}n/2$, where $g_{\rm ad}$ is the impurity-dimer s-wave coupling constant and $n/2$ is the dimer density. Since in the BEC limit,  $C_2/N=4\pi/a$,
identifying Eq. (\ref{eq:Born2}) with the mean-field impurity-dimer interaction  implies that

\be
F\left(\frac{1}{k_Fa}\right)\underset{a\rightarrow 0^+}{=}8\pi^2\kappa(\eta)\frac{m_{\rm f}}{m_{\rm r}}\frac{\ln\left(k_Fa\right)}{k_Fa}+...
\ee
 and

 \be
 g_{\rm ad}=2g'  \left[1-8\pi^2\kappa(\eta)\frac{m_{\rm f}}{m_{\rm r}}\frac{a'}{a}\left(\ln(R_3/a)+C_{\rm ad}\right)...\right],
    \label{eq:AtomDimer}
 \ee
where the constant $C_{\rm ad}$ can be obtained from the direct analysis of the atom-dimer scattering problem \cite{SuppMat}.

\begin{figure}
    \centering
    \includegraphics[width=\columnwidth]{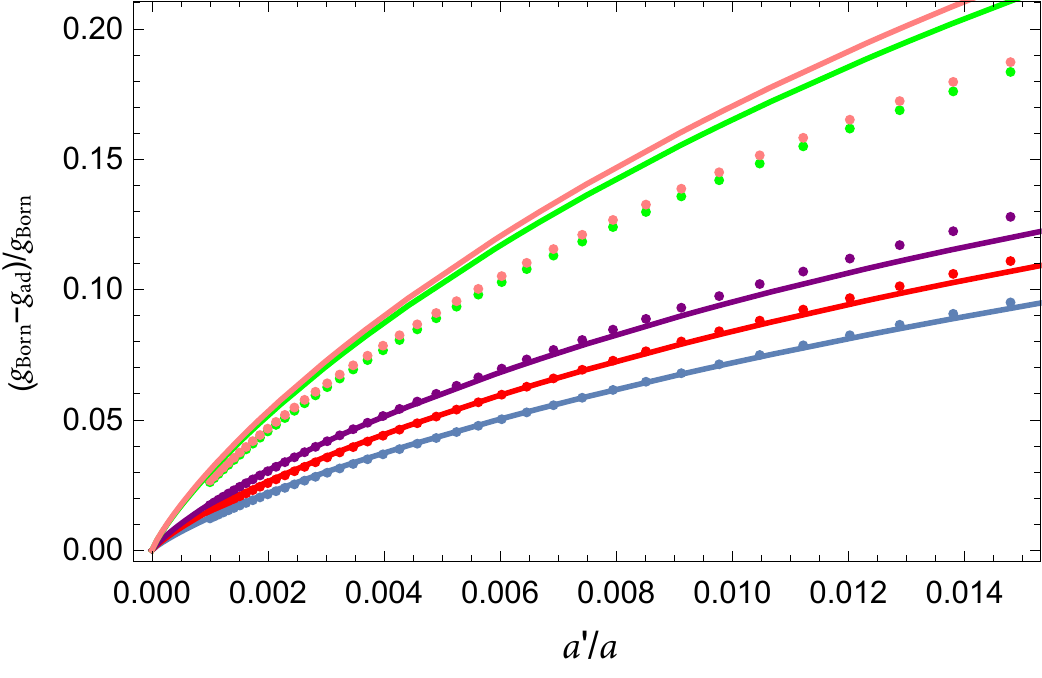}
    \caption{Atom-dimer s-wave coupling constant relatively to Born's approximation prediction $g_{\rm Born}=2g'$. Dots: numerical resolution of the three-body problem from \cite{zhang2014calibration}. From bottom to top: $\eta$= 7/40 (blue), 23/40 (red), 7/6 (purple), 87/6 (orange), 133/6 (green). Solid line: Asymptotic result Eq. (\ref{eq:AtomDimer}) where $R_3$ and $C_{\rm ad}$  are computed numerically and are given in \cite{SuppMat}. Here $R_3$ and $C_{\rm ad}$ are computed taking $R_e=0$ and the slight discrepancy observable at large $\eta$ is probably  due to the finite range used in \cite{zhang2014calibration} to regularize the three-body problem.}
    \label{fig:Fig5}
\end{figure}

Eq. (\ref{eq:Born2}) can be used to benchmark previous works on this problem. Ref. \cite{yi2015polarons,nishida2015polaronic} were based on a mean-field description of the superfluid component of the system.  The mean-field calculation obeys BCS and BEC asymptotic behaviours similar to those predicted by Eq. (\ref{eq:Born2}) except for the value of $\kappa$ that does not coincide with the present result since the last term in Eq. (\ref{eq:Kappa}) is missing within a BCS approach \cite{SuppMat}. This discrepancy is easily understandable. Indeed, this term corresponds to the third diagram of Fig. (\ref{fig:diagrams}) where the two fermions interacts between their interaction with the impurity, which contradicts the BCS assumption of non-interacting Bogoliubov excitations.  For $\eta=1$, $\kappa/\kappa_{\rm MF}\simeq 15$, showing that BCS approximation underestimates strongly beyond mean-field contributions.
Eq. (\ref{eq:AtomDimer}) can also be compared to the numerical calculation of the atom-dimer scattering length reported in \cite{zhang2014calibration}. The comparison between numerics and our analytical result for experimentally relevant mass ratios  is shown in Fig. (\ref{fig:Fig5}) which demonstrates a very good agreement between the two approaches. Note also that Eq. (\ref{eq:AtomDimer}) clarifies the range of validity of the perturbative expansion. In addition to the diluteness assumption $k_F |a'| \ll 1$, the validity of Born's expansion requires the additional condition $|a'|/a\ll 1$ when $a >0$.

Finally, the convergence of Eq. (\ref{Eq:MainEq}) entails that $\chi$ must obey the large momentum asymptotic behavior
\be
\chi(q,\varepsilon^{\rm (i)}_q)\underset{q\rightarrow\infty}{=}\frac{1}{\varepsilon^{\rm (r)}_q}\left[1-\pi^2\kappa(\eta)\frac{m_{\rm f}}{m_{\rm r}}\frac{C_2}{Nq}+...\right]
\label{Eq:AsymptoticChi}
\ee
For $m_{\rm i}\rightarrow\infty$, $\varepsilon^{(\rm i)}_q= 0$, we have $\kappa(\infty)=-1/4\pi$ and we recover the asymptotic result derived in \cite{hofmann2017deep} using operator product expansion.  Note that mean-field theory  predicts $\kappa_{\rm MF}(\infty)=0$, and therefore disagrees with this independent result.

Using the mean-field estimate for $F$ at unitarity, we see that the second-order correction to the polaron energy (Eq. (\ref{eq:Born2})) is dominated by the logarithmic contribution. In the case of the polaron oscillation experiments  reported in \cite{Ferrier2014Mixture}, the predicted correction amounts to a 5\% shift of the oscillation frequency. Although small, this correction is within the reach of current experimental capabilities and shows that the results presented in this work are necessary to achieve the percent-level agreement between experiment and theory targeted by state of the art precision quantum many-body physics.

\begin{acknowledgements}
The authors thank M. Parish, J. Levinsen, F. Werner, C. Mora, L. Pricoupenko as well as ENS ultracold Fermi group for insightful discussions. They thank Ren Zhang and Hui Zhai for providing the data shown in Fig. \ref{fig:Fig5}. This work was supported by ANR (SpifBox) and EU (ERC grant CritiSup2).
\end{acknowledgements}

\bibliographystyle{unsrt}
\bibliography{bibliographie}

\clearpage


\section*{Supplementary material}

# Few vs many-body physics of an impurity immersed in a superfluid of spin 1/2 attractive fermions: Supplemental information

M. Pierce,[1] X. Leyronas,[2] and F. Chevy[1]

[1]*Laboratoire Kastler Brossel, ENS-Université PSL, CNRS, Sorbonne Université, Collège de France.*
[2]*Laboratoire de physique de l'Ecole normale supérieure,
ENS, Université PSL, CNRS, Sorbonne Université,
Université Paris-Diderot, Sorbonne Paris Cité, Paris, France.*

## THREE-BODY PHASE-DIAGRAM

### Efimov ground state

The Efimov spectrum can be obtained without further regularization of the three-body problem using a two-channel model described by the hamiltonian [1, 2]

$$\hat{H} = \sum_{\mathbf{k},\sigma} \epsilon_k \hat{a}^\dagger_{\mathbf{k},\sigma} \hat{a}_{\mathbf{k},\sigma} + \sum_{\mathbf{K},\sigma} (E_{\sigma,0} + \epsilon_k/2) \hat{b}^\dagger_{\mathbf{K},\sigma} \hat{b}_{\mathbf{K},\sigma}$$
$$+ \frac{\hbar^2}{2m}\sqrt{\frac{2\pi}{R_e}} \sum_{\substack{\mathbf{K},\mathbf{k},k<\Lambda \\ \sigma_1 \neq \sigma_2 \neq \sigma_3}} (\hat{b}^\dagger_{\mathbf{K},\sigma_1} \hat{a}_{\mathbf{k}+\mathbf{K}/2,\sigma_2} \hat{a}_{-\mathbf{k}+\mathbf{K}/2,\sigma_3} + \text{H.c.}). \quad (S1)$$

In this expression, $\sigma \in \{1,2,3\}$ labels the three atomic species, $\hat{a}_{\mathbf{k},\sigma}$ is the atomic (open channel) annihilation operator for species $\sigma$, $\hat{b}_{\mathbf{K},\sigma}$ the molecular (closed channel) annihilation operator describing dimers not involving spin $\sigma$ atoms, and $R_e$ the effective range of the potential. For the sake of simplicity, we assume that all three atomic species have the same mass $m$ and that the atom-dimer coupling is the same for all three species. The atom-atom scattering lengths are nevertheless controlled independently by the bare molecular binding energies $E_{\sigma,0}$.

The solutions of the two-body problem shows that the scattering length $a_\sigma$ between two atoms $(\sigma_1,\sigma_2) \neq \sigma$ is given by

$$\frac{1}{a_\sigma} = \frac{2}{\pi}\Lambda - \frac{R_e m E_{\sigma,0}}{\hbar^2}, \quad (S2)$$

where $\Lambda$ is a UV momentum cut-off.

The three-body bound states are described by the Ansatz

$$|\psi\rangle = \sum_{\mathbf{k}_1,\mathbf{k}_2} \beta(\mathbf{k}_1,\mathbf{k}_2) \hat{a}^\dagger_{\mathbf{k}_1,1} \hat{a}^\dagger_{\mathbf{k}_2,2} \hat{a}^\dagger_{-\mathbf{k}_1-\mathbf{k}_2,3}|0\rangle$$
$$+ \sum_{\sigma,\mathbf{k}} \frac{\alpha_\sigma(\mathbf{k})}{k} \hat{a}^\dagger_{-\mathbf{k},\sigma} \hat{b}^\dagger_{\mathbf{k},\sigma}|0\rangle. \quad (S3)$$

where $|0\rangle$ is the vacuum.

The trimer energy $E_3 = -\hbar^2 \kappa^2/m$ is then obtained by

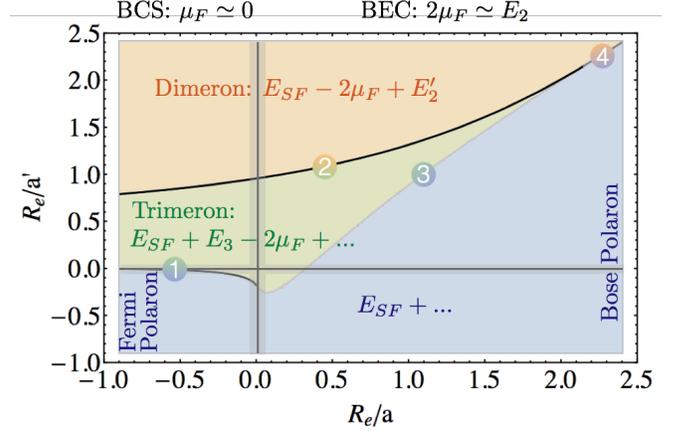

FIG. S1. Phase diagram of an impurity immersed in a fermionic superfluid. The typical energy of each phase (detailed in the text) is written in each domain, the dots in each expression contain the mean-field terms that are negligible compared to the other terms. The typical expression of the chemical potential in the BEC-BCS crossover is given at the top (on the BCS side, it corresponds approximately to the Fermi energy which is negligible compared to the other energy scales presented here, hence the 0). The four numbers correspond to the four frontiers described in the text. The grey bands correspond to the parameter range where many-body effects affect few-body physics for a typical value $k_F R_e = 5 \times 10^{-2}$.

solving the set of three equations

$$\left[\sqrt{1+3p^2/4} + \kappa R_e(1+3p^2/4) - \frac{1}{\kappa a_\sigma}\right] \alpha_\sigma(p)$$
$$= \frac{1}{\pi} \int_0^\infty dq \ln\left(\frac{p^2+q^2+pq+1}{p^2+q^2-pq+1}\right) \left[\sum_{\sigma'\neq\sigma} \alpha_{\sigma'}(q)\right], \quad (S4)$$

### Frontiers of the stability diagram

We explain here how the different domains of the stability diagram of Fig. S1 are obtained (corresponding to Fig. 2 of the main text).

In experiments, $R_e$ is small compared to the interparticle distance, therefore we have $R_e k_F \ll 1$, where the Fermi wavevector $k_F$ is defined by $n \equiv k_F^3/(3\pi^2)$. On the BCS side of the BEC-BCS crossover of the super-

fluid, we have $1/(k_F a) < 0$ and $1/|k_F a| \sim 1$ and therefore $|R_e/a| = R_e k_F/|k_F a| \ll 1$. For the same reason, the BEC side corresponds to $1/(k_F a) \sim 1$ and we have $R_e/a = R_e k_F/(k_F a) \ll 1$. On the graph of Fig. S1, the crossover region of the superfluid is therefore concentrated in a narrow region around the $y$-axis. The consequence of this separation of scales is that, except in this narrow region, for $R_e/a > 0$, the superfluid is made of weakly interacting tightly bound dimers, while for $R_e/a < 0$, it is made of extremely loose Cooper pairs corresponding essentially to non-interacting fermions (except for superfluid properties).

Concerning the impurity-fermion interaction regime, we have similarly $(k_F a') = (R_e/a')^{-1}(k_F R_e)$, which is a small dimensionless number for $R_e/a'$ not too small. Therefore, except in a narrow region around the $x$-axis in Fig. S1, we have $k_F |a'| \ll 1$.

In the $(R_e/a' < 0, R_e/a < 0)$ quadrant (Fermi polaron sector), the energy of the impurity immersed in the superfluid is given by $E_{SF} + g'n$, where $E_{SF}$ is the ground state energy of the superfluid without the impurity.

In the Bose-polaron sector $(R_e/a' < 0, R_e/a > 0)$, the energy of the polaron in the superfluid is given by $E_{SF} + g'_{b-f} n$, where $g'_{b-f}$ is the coupling constant between the impurity and a dimer of the superfluid.

In the $(R_e/a' > 0, R_e/a < 0)$ quadrant the energy of the dimeron in the superfluid is $E_{SF} - 2\mu_f + E'_2 + g'_{b-f} n + gn$. The $-2\mu_f$ contribution originates from Cooper pair breaking. One of the fermions of the pair binds to the impurity to form a dimer of energy $E'_2$ while the second remains unbound and contributes to the mean-field term $gn$. For $R_e/a < 0$, $\mu_f \approx \hbar^2 k_F^2/(2m)$ is the chemical potential of an ideal gas. The dimer energy is given by [3] $E'_2 = -\frac{\hbar^2}{m} \kappa'^2_2$, with $R_e \kappa'_2 = 2\frac{R_e}{a'}\left(1+\sqrt{1+4\frac{R_e}{a'}}\right)^{-1}$.

For $(R_e/a' > 0)$, and $(R_e/a > 0)$, the energy of the dimeron in the superfluid is $E_{SF} - E_2 + E'_2 + g'_{b-b} n + gn$ where we have substracted the dimer energy $E_2$ lost by the binding of a fermion of the bath with the impurity. The expression for $E_2$ is obtained from the expression of $E'_2$ by replacing $a'$ by $a$. $g'_{b-b}$ is the coupling constant characterizing the interaction between the dimeron and the dimers of the superfluid.

Finally, the energy of the trimeron in the superfluid is given by $E_{SF} + E_3 - 2\mu_f + g'_{t-f} n$. $E_3$ is the trimer energy (a 3-body bound state), $g_{t-f}$ is a trimer-fermion coupling constant, and $-2\mu_f$ is the energy of the two fermions of the trimer coming from the superfluid bath.

Depending on the parameters, we determine which state has the lowest energy. In this limit where $k_F$ tends to zero, all the terms containing $n$ or $k_F$ vanish. All these energies, minus the neglected mean-field terms, are gathered in Fig. S1. We then consider four cases (corresponding to the four numbers displayed in Fig. S1):

1. Fermi polaron $(R_e/a < 0)$ vs trimeron. The frontier is obtained by solving $E_3(R_e, a, a') = 0$.

2. Dimeron vs trimeron. The frontier is obtained by solving $E_3(R_e, a, a') = E'_2$.

3. Bose polaron $(R_e/a > 0)$ vs trimeron. The frontier is obtained by solving $E_3(R_e, a, a') = E_2$.

4. Bose polaron vs dimeron. The frontier is obtained by solving $E'_2 = E_2$. Since $E'_2$ and $E_2$ are given by the same function evaluated for $a'$ and $a$, this frontier is included in the line $a' = a$, i.e. the first bisector in S1.

We end this section by noticing that the Fermi polaron-trimeron frontier approaches the $x$-axis for $Re/a \to -\infty$. In this region, the calculation is no more controlled ($k_F$ is not negligible anymore). However, since the polaron/trimeron is a crossover [4], the transition line cannot be defined precisely anyway.

Moreover we note that for $a = a'$, our results agree with the calculations reported in [3] for three-component color Fermi gases.

## BCS THEORY

This section presents a derivation of the results presented in the paper within the simplified framework of BCS mean-field theory. We first determine $\chi$, then the $F$ function and finally we compare it to the exact expressions presented in this paper.

Since these results will only confirm the general behaviour but will not yield quantitative predictions, we restrain our calculations to the simplest case $m_f = m_i = m$ so for $\eta = 1$, a ratio close to the mass ratio we have with our Lithium experiment (7/6).

### Mean-field compressibility

As described in the main text, second order perturbation theory relates the polaron energy shift to the fermionic superfluid dynamical compressibility $\chi(\boldsymbol{q}, E)$ defined by

$$\chi(\boldsymbol{q}, E) = \frac{1}{N} \sum_\alpha \frac{|\langle \alpha | \hat{\rho}_{-\boldsymbol{q}} | 0 \rangle|^2}{E_\alpha - E_0 + E}. \quad (S5)$$

Here, we derive the expression of $\chi$ using BCS theory where the fermionic medium is described by the mean-field Hamiltonian

$$\hat{H}_{\mathrm{mb}} = \sum_{\boldsymbol{k},\sigma} \xi_k \hat{c}^\dagger_{\boldsymbol{k}\sigma} \hat{c}_{\boldsymbol{k}\sigma} + \Delta^* \sum_{\boldsymbol{k}} \hat{c}_{\boldsymbol{k}\uparrow} \hat{c}_{-\boldsymbol{k}\downarrow} + \mathrm{h.c.} \quad (S6)$$

with $\xi_k = \epsilon_k^{(\mathrm{f})} - \mu$, $\mu$ is the chemical potential, and the gap $\Delta$ is defined by

$$\Delta = \frac{g_0}{\Omega} \sum_{\boldsymbol{k}} \langle \hat{c}_{-\boldsymbol{k}\downarrow} \hat{c}_{\boldsymbol{k}\uparrow} \rangle. \tag{S7}$$

The Hamiltonian is diagonalized by introducing Bogoliubov operators $\hat{\gamma}_{\boldsymbol{k}\pm}$ defined by:

$$\hat{c}_{\boldsymbol{k}\uparrow} = u_k \hat{\gamma}_{\boldsymbol{k}+} - v_k \hat{\gamma}_{-\boldsymbol{k}-} \tag{S8}$$
$$\hat{c}_{\boldsymbol{k}\downarrow} = u_k \hat{\gamma}_{\boldsymbol{k}-} + v_k \hat{\gamma}_{-\boldsymbol{k}+}, \tag{S9}$$

with

$$u_k = \sqrt{\frac{1}{2}\left(1 + \frac{\xi_k}{E_k}\right)}, \quad v_k = \sqrt{\frac{1}{2}\left(1 - \frac{\xi_k}{E_k}\right)} \tag{S10}$$

and $E_k = \sqrt{\xi_k^2 + |\Delta|^2}$.

To derive the BCS expression of the compressibility, we express the matrix elements $\langle \alpha | \hat{\rho}_{-\boldsymbol{q}} | 0 \rangle = \sum_\sigma \langle \alpha | \hat{c}^\dagger_{\boldsymbol{k}-\boldsymbol{q},\sigma} \hat{c}_{\boldsymbol{k},\sigma} | 0 \rangle$ using the Bogoliubov creation and annihilation operators. After a straightforward calculation, we finally obtain

$$\chi^{\mathrm{MF}}(\boldsymbol{q}, E) = \frac{1}{N} \sum_{\boldsymbol{k}} \frac{2 u_{k-q}^2 v_k^2 + 2 u_k v_k u_{k-q} v_{k-q}}{E_k + E_{k-q} + E}, \tag{S11}$$

where we have used the fact that the excited states $|\alpha\rangle$ correspond to pairs of Bogoliubov excitations, hence $E_\alpha - E_0 = E_k + E_{k-q}$. We use the notation MF to signify that this result is only valid in BCS theory, a mean-field theory.

**Perturbative calculation of the energy**

In order to calculate the polaron energy shift, we need to consider the perturbative development we obtained in the article, adapted to BCS theory:

$$\Delta E_{\mathrm{pert}}^{\mathrm{MF}} = \left[ g'n + \frac{g'^2 n}{\Omega} \sum_{\boldsymbol{q}} \left( \frac{1}{\varepsilon_q^{(\mathrm{r})}} - \chi^{\mathrm{MF}}(\boldsymbol{q}, \varepsilon_q^{(\mathrm{i})}) \right) \right] \tag{S12}$$

After turning sums to integrals and performing the angular integrations we can write this expression as:

$$\Delta E_{\mathrm{pert}}^{\mathrm{MF}} = g'n + \frac{g'^2}{8\pi^4} \frac{m}{\hbar^2} \int k^2 \mathrm{d}k \int q^2 \mathrm{d}q \Bigg( \frac{4 v_k^2}{q^2} - \frac{2 u_q^2 v_k^2 + 2 u_k v_k u_q v_q}{kq} \ln\left( \frac{E_k + E_q + \frac{\hbar^2(k+q)^2}{2m}}{E_k + E_q + \frac{\hbar^2(k-q)^2}{2m}} \right) \Bigg) \tag{S13}$$

To study the behaviour of these integrals for high $k$, we perform the variable change $k \to u = (k/k_F)/\sqrt{|\Delta|/E_F}$, $q \to v = (q/k_F)/\sqrt{|\Delta|/E_F}$ and we get

$$\Delta E_{\mathrm{pert}}^{\mathrm{MF}} = g'n \left[ 1 + k_F a' \frac{3}{2\pi} \left| \frac{\Delta}{E_F} \right|^2 I(\Lambda/k_F) \right] \tag{S14}$$

with $I$ corresponding to the integral left to calculate in Eq. (S13) that depends on the cut-off $\Lambda$ and also on the ratio $\mu/|\Delta|$.

In the limit $u, v \gg 1$, we can simplify greatly the expression of the integral $I$.

First, we can see that the terms $u_k^2$ and $v_k^2$ can be rewritten, in this limit:

$$u_k^2 \sim 1, \quad v_k^2 \sim \frac{1}{4} \frac{|\Delta|^2}{\xi_k^2} \to \frac{1}{4u^4} \tag{S15}$$

From this last expression, we can also get Tan's contact for two fermions $C_2$ in BCS theory. Indeed, using the property of momentum distribution [5]:

$$n_\uparrow(k) \underset{k\to\infty}{\sim} n_\downarrow(k) \underset{k\to\infty}{\sim} \frac{C_2}{k^4} \tag{S16}$$

and knowing that in BCS theory we have $n(k) = n_\uparrow(k) + n_\downarrow(k) = 2 v_k^2 \Omega$, we see that we have the right dependence for the momentum distribution and we can extract the contact

$$\frac{C_2}{N} = \frac{3\pi^2}{4} \left| \frac{\Delta}{E_F} \right|^2 k_F. \tag{S17}$$

The integral can then be simplified in this limit as

$$I(\Lambda/k_F) = \int \frac{\mathrm{d}u}{u} \int \frac{\mathrm{d}v}{u} \Bigg[ 1 - \frac{1}{2}\left(\frac{v}{u} + \frac{u}{v}\right) \ln\left(\frac{1 + v/u + (v/u)^2}{1 - v/u + (v/u)^2}\right) \Bigg]. \tag{S18}$$

The second integral (over $v/u$) converges towards $2\pi^4 \kappa^{\mathrm{MF}}$ and the first integral (over $u$) gives the logarithmic divergence. We can finally write:

$$I = 2\pi^4 \kappa^{\mathrm{MF}} (\ln(\Lambda/k_F) + ...) \tag{S19}$$

with $\kappa^{\mathrm{MF}}$:

$$\kappa^{\mathrm{MF}} = \frac{\sqrt{3}}{8\pi^3} - \frac{1}{12\pi^2}. \tag{S20}$$

In the main text we found:

$$\kappa(1) = \frac{\sqrt{3}}{8\pi^3} - \frac{1}{12\pi^2} - \frac{1}{9\pi\sqrt{3}} \tag{S21}$$

The two results are very similar except for the last term of $\kappa(1)$ which does not appear in the mean-field approach because BCS theory does not account for interactions between excitations of the superfluid. This missing term is actually pretty important since it is the leading term in $\kappa$, therefore we get a ratio $\kappa(1)/\kappa^{\mathrm{MF}} \simeq 15$.

### The $F$ function

We find out analytically that there is again a logarithmic divergence of this second order term, consistently with what we stated before. By combining equations (S14), (S17) and (S19), we get the expression of the energy calculated up to second order in perturbation using BCS theory :

$$\Delta E^{\mathrm{MF}}_{\mathrm{pert}} = g'n \left[ 1 + k_F a' F^{\mathrm{MF}}\left(\frac{1}{k_F a}\right) \right. \tag{S22}$$
$$\left. + 4\pi \kappa^{\mathrm{MF}} a' \frac{C_2}{N} \ln(\Lambda/k_F) \right]$$

with $F^{\mathrm{MF}}$ a function that can be computed numerically throughout the BEC-BCS crossover by calculating the difference between the exact expression of the integral $I$ defined in Eq. (S13) and the logarithmic term we obtained in Eq. (S19). These numerical calculations show that this function does not depend on the cut-off but only on the parameter $1/(k_F a)$.

Then, by introducing a similar renormalization with a three-body term, we can rewrite the energy as:

$$\Delta E^{\mathrm{MF}} = g'n \left[ 1 + k_F a' F^{\mathrm{MF}}\left(\frac{1}{k_F a}\right) \right.$$
$$\left. -4\pi \kappa^{\mathrm{MF}} \frac{a' C_2}{N} \ln(k_F R_3) + ... \right], \tag{S23}$$

We get a very similar expression to the one we found in this letter, only we replaced $F$ and $\kappa(1)$ by $F^{\mathrm{MF}}$ and $\kappa^{\mathrm{MF}}$.

The function $F^{\mathrm{MF}}$ is represented in Fig. S2, and we can observe the two asymptotic behaviours on the BCS and BEC sides:

1. In the BCS limit we recover once again the Fermi-polaron, hence $F^{\mathrm{MF}}(-\infty) = 3/2\pi$ for $\eta = 1$.

2. In the BEC limit, we get a behaviour consistent with the Bose-Polaron:

$$F^{\mathrm{MF}}\left(\frac{1}{k_F a}\right) = 16\pi^2 \kappa^{\mathrm{MF}} \frac{\ln(k_F a)}{k_F a} + ... \tag{S24}$$

In conclusion, BCS theory predicts the correct qualitative behaviour for the polaron energy shift but is quantitatively wrong, which is illustrated in Fig. S5 at the end of this supplementary material.

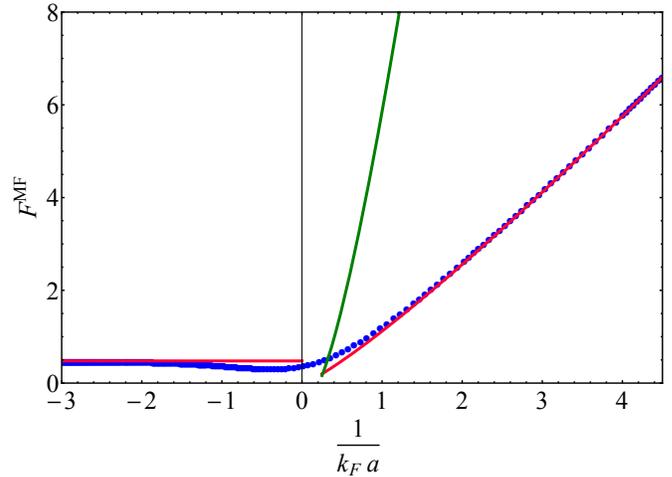

FIG. S2. Blue dots: Representation of the function $F^{\mathrm{MF}}$ through the crossover for $\eta = 1$. Red curves: Asymptotic behaviours described in the text on the BCS side: $F^{\mathrm{MF}}(-\infty) = \frac{3}{2\pi}$, and the BEC Side : $F^{\mathrm{MF}}(X) \simeq 16\pi^2 \kappa^{\mathrm{MF}} X \ln(1/X) + A_0 X$ with $A_0$ an adjustable parameter found out to be, after optimization, $A_0 \simeq 1.1$. Green curve: asymptotic behaviour on the BEC side using the true value of $\kappa$, $C_{\mathrm{ad}}$ and $R_3$ (the last two are given in the last section of this Supplementary material).

## THREE-BODY PARAMETERS

### Calculating $R_3$

We can obtain the three-body parameter $R_3$ introduced in the equation

$$\Gamma_{\mathrm{Born}} - \Gamma_{\mathrm{Faddeev}} \underset{\Lambda \to \infty}{=} g'^2 \kappa(\eta) \ln(\Lambda R_3) + o(1) \tag{S25}$$

by calculating numerically this difference. We break down this term into three parts, each corresponding to one of the first three diagram of Fig. 3 from this letter.

Firstly, in the effective range approximation, the two-body T-matrix is given by

$$\hat{t}_i = \frac{g'/\Omega}{1 + ika' + R_e a' k^2} \tag{S26}$$

where $R_e$ is the effective range of the potential and $k$ is given by:

$$k = \sqrt{2m_r(E + i0^+)/\hbar^2} \tag{S27}$$

with $E$ the energy of the initial state in the center-of-mass frame of the three particles at the moment of the interaction. For the diagrams we want to calculate, we only have to consider the case where the fermions have impulsions of $\boldsymbol{p}$ and $-\boldsymbol{p}$ and the impurity has an impulsion equal to zero (cf Fig. 3 from the article). This leads to:

$$E/\hbar = 0 - \left(\frac{p^2}{2m_f} + \frac{p^2}{2(m_f+m_b)}\right) = -\frac{3}{4}\frac{p^2}{m} \quad (S28)$$

Finally, the two-body T-matrix can be written as:

$$\hat{t}_i = \frac{g'/\Omega}{1-\sqrt{\frac{\eta(2+\eta)}{(1+\eta)^2}}a'p - \frac{\eta(2+\eta)}{(1+\eta)^2}\left(\frac{R_e}{a'}\right)(a'p)^2 + i0^+} = \frac{g'}{\Omega}t(p) \quad (S29)$$

Then, we can write below the expressions corresponding to each of these three diagrams.

$$\Gamma^{(1)} = -2\frac{m_f^2}{\hbar^4}\frac{g'}{\Omega}\sum_{\mathbf{p}}\frac{1}{p^4}t(p) \quad (S30)$$

$$\Gamma^{(2)} = -2\frac{m_f^3}{\hbar^6}\frac{g'^2}{\Omega^2}\sum_{\mathbf{p_1},\mathbf{p_2}}\frac{1}{p_1^2 p_2^2}\frac{t(p_1)t(p_2)}{(p_1^2+p_2^2)(\frac{\eta+1}{2\eta}) - \frac{1}{\eta}\vec{p_1}\cdot\vec{p_2}} \quad (S31)$$

$$\Gamma^{(3)} = -4\frac{m_f^3}{\hbar^6}\frac{g'^2}{\Omega^3}\sum_{\mathbf{p_1},\mathbf{p_2},\mathbf{p_3}}\left[\frac{1}{p_1^2 p_3^2}\frac{4\pi}{1/a - p_2\sqrt{\frac{\eta+2}{4\eta}}}\right.$$
$$\left.\times \frac{t(p_1)}{p_1^2 + p_2^2(\frac{\eta+1}{2\eta}) - \vec{p_1}\cdot\vec{p_2}}\frac{t(p_3)}{p_3^2 + p_2^2(\frac{\eta+1}{2\eta}) - \vec{p_3}\cdot\vec{p_2}}\right] \quad (S32)$$

In order to calculate the Faddeev term for $\Gamma$, one has to use the expression of $t(p)$ given in S29. On the other hand, to obtain the Born term, one has to expand this expression of $t(p)$ up to first order in $a'$ for $\Gamma^{(1)}$ and up to zero order for the other two components (so just replacing it by 1), so that all three components of $\Gamma$ are expanded up to order two in $a'$.

For $\Gamma^{(3)}$, we calculate the sum in the limit $1/a \ll 1/a'$ (highly interacting fermions) in which the difference between the Faddeev term and the Born term does not depend on $a$. To calculate these different sums we proceed similarly as we did in the previous section for BCS theory.

In this framework, one can show that $R_3/a'$ only depends on the ratio $R_e/a'$ and the mass ratio. We show in Fig. S3 the numerical calculations of the difference $\Gamma_{\text{Born}} - \Gamma_{\text{Faddeev}}$ for the mass ratio $\eta = 7/6$ and $R_e/|a'| = 1$. We see that we indeed get the logarithmic behaviour with $\kappa(7/6)$ as the proportionality constant.

We show in Fig. S4 the parameter $R_3/|a'|$ for different values of the ratio $R_e/|a'|$ in the case $\eta = 7/6$. For $R_e/|a'| \ll 1$, we get the asymptotic behaviour $R_3 \simeq 1.50|a'|$. For $R_e/|a'| \gg 1$, we see that $R_3$ increases exponentially:

$$R_3 \underset{\frac{R_e}{|a'|}\gg 1}{\propto} \sqrt{R_e|a'|}\exp\left(\frac{\sqrt{3}}{16\pi^2|\kappa(7/6)|}\sqrt{\frac{R_e}{|a'|}}\right). \quad (S33)$$

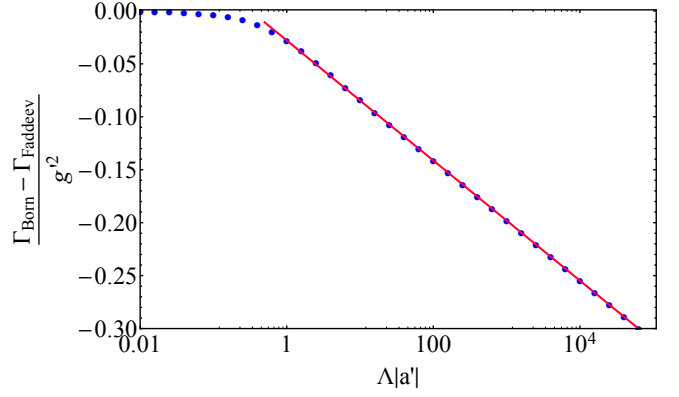

FIG. S3. Blue dots: numerical calculations of the left-hand side of eq. (S25), divided by $g'^2$, for $\eta = 7/6$ and $R_e = |a'|$. Red curve: fitting curve of the blue dots in the limit $\Lambda|a'| \gg 1$. We fit the data for $\Lambda|a'| \gg 1$ with the function $\kappa(7/6)\ln(X \times A_0)$ with $A_0$ a fitting parameter. The parameter $A_0$ gives us the value of $R_3/|a'|$: here we get $A_0 \simeq 3.10$.

At this point we should remind that we consider expansions for $\Lambda|a'| \ll 1$ but with $R_e/|a'|$ as an independent parameter with a given value. Consequently, we consider this exponential term as a constant included in $R_3$ in our perturbative calculations.

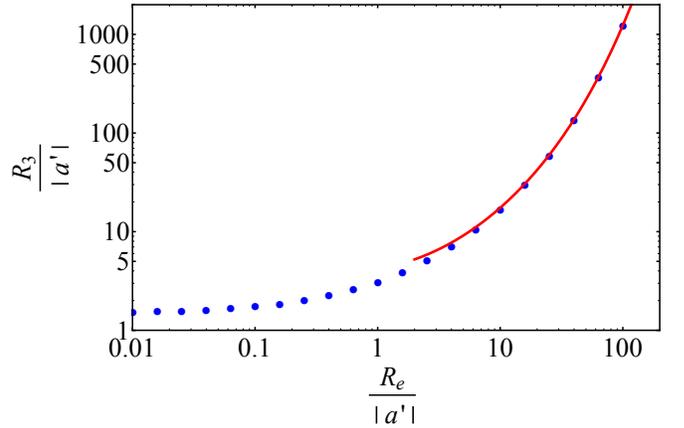

FIG. S4. Blue dots: numerical calculations of $R_3/|a'|$ for different $R_e/|a'|$ ratios and $\eta = 7/6$. Red curve: fit for $R_e/|a'| \gg 1$ using a function $A\sqrt{X}\exp\left(\frac{\sqrt{3}}{16\pi^2|\kappa(7/6)|}\sqrt{X}\right)$, with $A$ an adjustable parameter. $A \simeq 0.8$ after optimization.

To see the dependence on the mass ratio $\eta$, Table I lists numerical values of the parameter $R_3$ that were computed for experimentally relevant mass ratios and $R_e = 0$.

| $\eta$ | 7/40 | 23/40 | 7/6 | 87/6 | 133/6 |
|---|---|---|---|---|---|
| $R_3/a'$ | 1.03 | 1.41 | 1.50 | 1.46 | 1.46 |

TABLE I. Dimensionless parameter characterizing the Born expansion of the three-body scattering amplitude (Eq. (S25)) for $R_e = 0$ .

### Atom-dimer scattering

The atom-dimer T-matrix can be computed using the same approach. Indeed, since the fermions are asymptotically bound, we can treat the impurity-fermion interaction as a perturbation. This leads to the same diagrams as in the three-body scattering problem and the atom-dimer scattering length consequently suffers from the same logarithmic divergence when the range of the potential vanishes. For large $\Lambda$ the associated $T$-matrix scales as

$$T_{\rm ad}^{(1)} = \frac{2g'}{\Omega} \left[ 1 + 8\pi^2 \frac{m_f}{m_r} \kappa(\eta) \frac{a'}{a} \left( \ln(\Lambda a) + C_{\rm ad} + ... \right) \right] \quad (S34)$$

where the constant $C_{\rm ad}$ is computed numerically and is given in Table II for experimentally relevant values of the impurity-fermion mass ratios.

| $\eta$ | 7/40 | 23/40 | 7/6 | 87/6 | 133/6 |
|---|---|---|---|---|---|
| $C_{\rm ad}$ | 1.52 | 1.59 | 1.56 | 1.37 | 1.36 |

TABLE II. Dimensionless parameter characterizing the Born expansion of the atom-dimer scattering amplitude (Eq. (S34)) for $R_e = 0$.

The logarithmic divergence is once again cured by introducing the three-body interaction. Using the renormalized expression of $g_3(\Lambda)$ the three-body interaction contribution to the atom-dimer $T$-matrix amounts to

$$T_{\rm ad}^{(2)} = -\frac{16\pi^2 g'}{\Omega} \frac{m_f}{m_r} \kappa(\eta) \frac{a'}{a} \ln(\Lambda R_3). \quad (S35)$$

We indeed recover the asymptotic result Eq. [20] from the main text since we have

$$T_{\rm ad} = T_{\rm ad}^{(1)} + T_{\rm ad}^{(2)} =$$
$$T_{\rm ad,Born} \left[ 1 - 8\pi^2 \frac{m_f}{m_r} \kappa(\eta) \frac{a'}{a} \left( \ln(R_3/a) + C_{\rm ad} + ... \right) \right] \quad (S36)$$

where $T_{\rm ad,Born} = 2g'/\Omega$ corresponds to an atom-dimer scattering length $a_{\rm ad,Born}/a' = 4(1+\eta)/(2+\eta)$.

Finally, to highlight the shortcomings of BCS theory and the consistency of our three-body calculations, we fit the atom-dimer scattering length calculated in [6], see Fig. S5. There is no hesitation possible in seeing that the coefficient before the log obtained through BCS theory (including $\kappa^{\rm MF}$) is too small to show the logarithmic behaviour whereas the real $\kappa$ enables a much better fit.

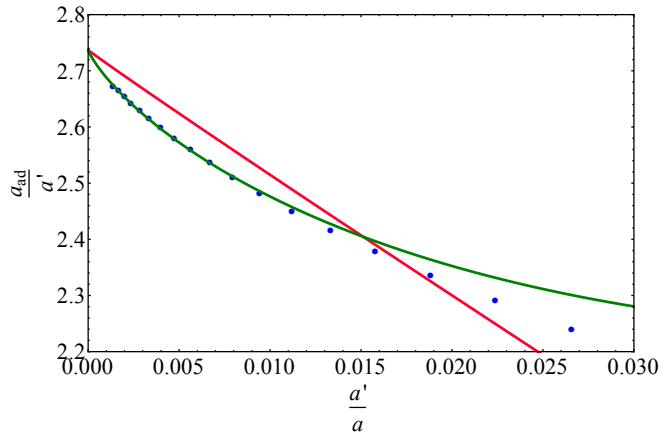

FIG. S5. Blue dots: Points from [6] showing the ratio of the atom-dimer scattering length $a_{\rm ad}$ over $a'$, for $\eta = 7/6$. Red solid curve: fit to the blue dots using the function $a_{\rm ad,\,Born}/a'(1 + AX(\ln X + B))$ where $A$ is a fixed parameter corresponding to the analytical result obtained through BCS theory ($A \propto \kappa^{\rm MF}$) and $B$ is an adjustable parameter. Green solid curve: theoretical curve obtained through three-body calculations, its equation is the same as the one used for the red curve but now with $A \propto \kappa$ and $B$ obtained through $C_{\rm ad}$ and $R_3$. We see that the curve corresponding to BCS theory (red) does not match at all the results reported in [6], contrary to the other one (green).

---



\end{document}